**Thermodynamic calculations using reverse Monte Carlo: Simultaneously tuning multiple short-range order parameters for 2D lattice adsorption problem**

Suhail Haque and Abhijit Chatterjee*

Department of Chemical Engineering, Indian Institute of Technology Bombay, Mumbai 400076, India

*Email: **abhijit@che.iitb.ac.in**

**Abstract**

Lattice simulations are an important class of problems in crystalline solids, surface science, alloys, adsorption, absorption, separation, catalysis, to name a few. We describe a fast computational method for performing lattice thermodynamic calculations that is based on the use of the reverse Monte Carlo (RMC) technique and multiple short-range order (SRO) parameters. The approach is comparable in accuracy to the Metropolis Monte Carlo (MC) method. The equilibrium configuration is determined in 5-10 Newton-Raphson iterations by solving a system of coupled nonlinear algebraic flux equations. This makes the RMC-based method computationally more efficient than MC, given that MC typically requires sampling of millions of configurations. The technique is applied to the interacting 2D adsorption problem. Unlike grand canonical MC, RMC is found to be adept at tackling geometric frustration, as it is able to quickly and correctly provide the ordered c(2×2) adlayer configuration for Cl adsorbed on a Cu (100) surface.

## 1. Introduction

The reverse Monte Carlo (RMC) method is used widely in the materials community for structural modelling[1,2]. The goal usually is to create a detailed atomistic configuration that is consistent with experimental structural data, such as x-ray or neutron diffraction data[3–15]. The advantage of solving such an inverse problem is that once a configuration is available, it can be analyzed to glean additional information about the structure, which may be inaccessible to the experiment. The input structural data contains short-range ordering information, e.g., the radial distribution function. Recently, our group has extended the application of the RMC method to a different problem; one which belongs to the area of lattice thermodynamic property calculations[16–19]. The main objective is to identify the short-range order (SRO) parameter at equilibrium[20] given the material interactions, composition and temperature as inputs. This quantity is subsequently used to generate equilibrium configurations using RMC. With the complete configurational details now being available[a], one can proceed to calculate thermodynamic properties, e.g., chemical potentials. Note that no experimental inputs are required in our approach. A similar philosophy is used here.

Free energy calculations are the starting point for prediction of phase diagrams [20–22], and for determining the direction in which thermodynamic processes like adsorption, absorption, segregation, aggregation, formation of ordered structures, catalytic reactions and phase transitions may proceed. As discussed later, from a computational efficiency standpoint RMC can be an attractive alternative to Monte Carlo (MC) based free energy calculations[23] while studying multicomponent systems. In our previous implementation, a detailed balance equation in terms one SRO parameter was written[17,19]. The detailed balance equation is a nonlinear algebraic equation. Finding its root yields the equilibrium SRO parameter. However, an issue is that obtaining a unique solution for multiple SRO parameters with one detailed balance equation is not possible. In this paper, we introduce a modified approach. Briefly, in the new methodology $m$ different SRO parameters are selected that can include pair, triplet, quadruplet probabilities and so on. A quantity called the flux is numerically evaluated for each SRO parameter. The flux provides a measure of the tendency of the system to stay at the current parameter value. The condition where the flux becomes zero is numerically determined. Details such as how the flux is to be evaluated, whether this approach can successfully work with multiple SRO parameters, and insights gained in the process, make this study distinct from our previous work.

---

[a] The underlying assumption here is that SRO parameters are accurate descriptors for the material structure.

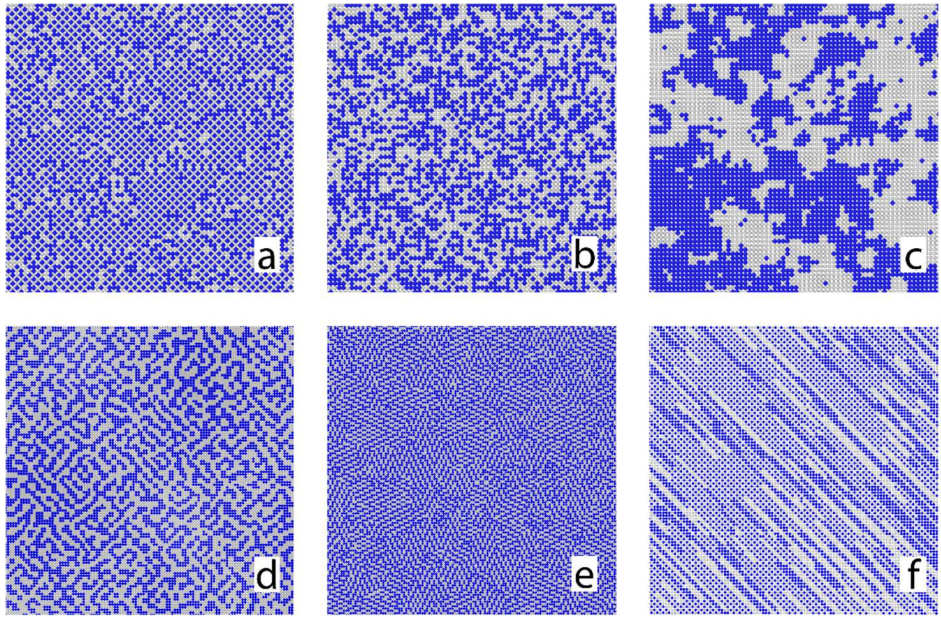

Figure 1: 2D configurations obtained for an adsorption model using the first nearest neighbor pair probability $z_{1NN}$ being ($a$) 0.2, ($b$) 0.5 and ($c$) 0.8 (see text for discussion on $z_{1NN}$). Panels

$(d) - (f)$ are generated using multiple SRO parameters (see Appendix). For (e) the fraction of filled sites is $x = 0.6$, whereas, for remaining panels $x = 0.5$.

RMC requires only the composition and SRO parameters as input. In certain situations, a single SRO parameter may be sufficient for explaining the local ordering behavior of atoms. Figure 1a-c shows examples of 2D configurations obtained with a single SRO parameter for the same composition. The configurations can be characterized as ordered structures (blue particles in a square arrangement in panel a) to a perfect well mixed arrangement (panel b) to one exhibiting phase separation and cluster formation (panel c). Figure 1d-f shows some exotic configurations that can be achieved only by specifying additional SRO parameters. Although such configurations may be rarely seen in nature, the point is that the use of multiple SRO parameters provides access to a wider configurational space. Later, we will provide examples that require multiple SRO parameters.

Section 2 expands on our discussion on the SRO parameter as well as the new RMC approach. In section 3, simulation details are provided. Examples illustrating application of RMC are presented in section 4. Finally, conclusions and salient features of our method are discussed in section 5.

## 2. Theory

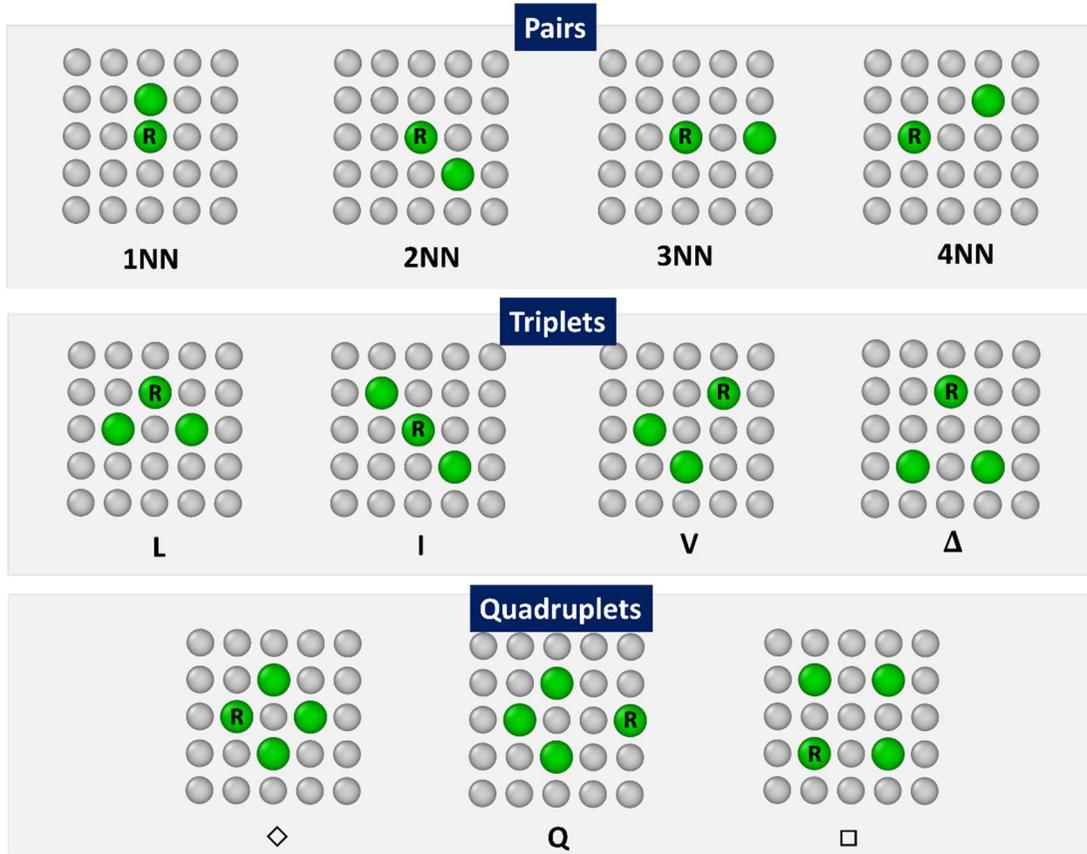

Figure 2: Site clusters in our cluster expansion model (Equation (1)). Sites belonging to the cluster are shown in green. Green indicates that the site is filled. Same clusters are used later as SRO parameters.

### 2.1 Interaction model

An adsorption model is used to illustrate our approach. Adsorption proceeds on a 2D square lattice. Consider a binary system $A_xV_{1-x}$, where $x$ represents the fraction of lattice sites occupied by $A$ atom and $V$ denotes vacant sites. $\sigma_i$ is the occupation at site $i$ ($\sigma_i = 1$ implies occupied site, otherwise $\sigma_i = 0$). The Hamiltonian for the system is given in the form of a cluster expansion

$$H(\{\sigma\}) = \sum_{\langle i,j \rangle} w_{ij}\sigma_i\sigma_j + \sum_{\langle i,j,k \rangle} w_{ijk}\sigma_i\sigma_j\sigma_k + \sum_{\langle i,j,k,m \rangle} w_{ijkm}\sigma_i\sigma_j\sigma_k\sigma_m + \cdots \quad (1)$$

Here, $w_{ij}$ is the interaction between a pair of atoms at sites $i$ and $j$. The interaction strength can depend on the separation, such as first nearest neighbor (1NN), second (2NN) and so on. Examples of interactions considered are shown in Figure 2. $w_{ij}$ can be obtained from *ab initio* electronic structure calculations (see section 4.3) or using empirical potentials. Similarly, there can be different triplet and quadruplet interactions (Figure 2). $\langle i,j \rangle$, $\langle i,j,k \rangle$, $\langle i,j,k,m \rangle$ in Equation (1) implies that all site pairs, triplets and quadruplets, respectively, are counted only once in the sum. The green colored circles in Figure 2 indicates that the sites need to be occupied for an interaction to be present. An equivalent way of writing Equation (1) involves cluster counts. The cluster count depends on the relative arrangement of the adsorbed species (see Figure 1 for example).

## 2.2 Short-range order parameter

$A$ and $V$ species may order/arrange themselves in a manner that is far from perfectly random. The SRO parameter is a useful mathematical quantity for understanding the local atomic structure. SRO refers to the likelihood of finding atoms of a particular type, in the vicinity of other atoms of same/another type. For instance, the arrangement of atoms around the site labelled $R$ in Figure 2 may be of interest. As mentioned earlier, RMC can in principle generate any configuration consistent with the specified probability.

One would expect an SRO parameter to lie between 0 and 1. In reality, topological constraints introduced by the lattice lead to a narrower range of allowed probabilities. For simplifying our discussion, the SRO parameters selected are analogous to the clusters considered for the interaction model (see Figure 2), i.e., 11 different parameters are chosen. As more parameters are specified, it constrains the allowed probabilities for other SRO parameters to a very narrow range. In effect, only few SRO parameters are independent and act as descriptors for the material structure. Other SRO parameters are dictated by the independent ones. For reasons of computational efficiency, we tune only a minimum number of SRO parameters.

We begin by considering pair SRO parameters, which are related to radial distribution function (rdf)[18]. Suppose the site marked as $R$ in Figure 2 is occupied, while the second site may or may not be occupied. Let $N_{nNN}$ denote the number of $A - A$ pairs formed at the $n^{th}$ nearest neighbor ($nNN$) position. The maximum number of $A - A$ pairs is obtained when $x = 1$, such that $N_{nNN} = \frac{1}{2} N_A c_n$. Here $c_n$ is the coordination number. The factor of $\frac{1}{2}$ prevents double counting. We define the SRO parameter for the $nNN$ position as

$$z_{nNN} = \frac{N_{nNN}}{\frac{1}{2} N_A c_n}. \tag{2}$$

For well mixed arrangement, it can be shown that $z_{nNN} = x$. In general, $z_{nNN}$ can be different from $x$ (as already seen in Figure 1).

Similarly, for a triplet and a quadruplet we require the site labelled $R$ in Figure 2 to be occupied. $z_{cluster}$ is the probability that the remaining sites in the cluster are also occupied. Analogous to the pair case, the SRO parameter can be evaluated using the general expression

$$z_{cluster} = \frac{N_{cluster}}{bN_A}. \qquad (3)$$

Here $N_{cluster}$ is the number of times a cluster appears in a RMC configuration. Table 1 provides the value for $b$. Alternate definitions of SRO parameters are allowed. The relation between $z_{cluster}$ and $x$ is also provided for the special case well mixed arrangement. In this case, $x$ dictates all SRO parameters, i.e., there are no independent SRO parameters.

**Table 1: Definition of SRO parameter for a 2D square lattice. See Figure 2 for meaning of the site clusters. The general form and the special case of well-mixed arrangement are shown.**

| SRO parameter for site cluster | General form | Well mixed arrangement |
|---|---|---|
| 1NN pair | $z_{1NN} = \dfrac{N_{1NN}}{2N_A}$ | $z_{1NN} = x$ |
| 2NN pair | $z_{2NN} = \dfrac{N_{2NN}}{2N_A}$ | $z_{2NN} = x$ |
| 3NN pair | $z_{3NN} = \dfrac{N_{3NN}}{2N_A}$ | $z_{3NN} = x$ |
| 4NN pair | $z_{4NN} = \dfrac{N_{4NN}}{4N_A}$ | $z_{4NN} = x$ |
| L Triplet | $z_L = \dfrac{N_L}{4N_A}$ | $z_L = x^2$ |
| I Triplet | $z_I = \dfrac{N_I}{2N_A}$ | $z_I = x^2$ |
| V Triplet | $z_V = \dfrac{N_V}{4N_A}$ | $z_V = x^2$ |

| Δ Triplet | $z_\Delta = \dfrac{N_\Delta}{4N_A}$ | $z_\Delta = x^2$ |
| Quadruplet ◇ | $z_\diamond = \dfrac{N_\diamond}{N_A}$ | $z_\diamond = x^3$ |
| Quadruplet $Q$ | $z_Q = \dfrac{N_Q}{4N_A}$ | $z_Q = x^3$ |
| Quadruplet □ | $z_\square = \dfrac{N_\square}{N_A}$ | $z_\square = x^3$ |

## 2.3 Flux calculation

The equilibrium SRO parameter is determined on the basis of the system interactions, composition $x$ and temperature $T$. Suppose a list of tunable SRO parameters $\mathbf{z}$ is specified, using RMC one can generate a corresponding 2D configuration. The problem reduces to finding whether this configuration is an equilibrium one. Answering this question also provides the direction to move in the $\mathbf{z}$-space that will take us to the equilibrium SRO parameter $\mathbf{z}_{eq}$. This is achieved by writing a population balance for each of the SRO parameters of interest

$$\frac{dN_{cluster}}{dt} = F_{cluster}(\mathbf{z}). \qquad (4)$$

Here $t$ is a fictitious time, $F_{cluster}(\mathbf{z})$ is the flux calculated for the RMC configuration, which measures the tendency of the system to move away from the current $\mathbf{z}$. In section 4, the number of tunable SRO parameters lies between $m = 1 - 3$, e.g., $\mathbf{z} = \{z_{1NN}, z_{2NN}, z_{3NN}\}$. The behavior of few more SRO parameters, e.g., $\tilde{\mathbf{z}} = \{z_{4NN}, z_L, z_I, z_V, z_\Delta, z_\diamond, z_Q, z_\square\}$, is also of interest to us. $\tilde{\mathbf{z}}$ is determined from the 2D configuration generated at $\mathbf{z}$. We shall soon see that the flux can be evaluated for all 11 SRO parameters. One can make Equation (4) independent of system size by dividing the left- and right-hand sides by the quantity $bN_A$ – this gives us the SRO evolution equation. At equilibrium, the flux is zero for all SRO parameters, i.e.,

$$F_{cluster}(\mathbf{z}_{eq}) = 0. \tag{5}$$

Before we describe how the flux is to be numerically evaluated, there are few points to discuss about Equation (5). Figure 3a provides a flowchart of the RMC based procedure. To update the $\mathbf{z}$, a subset of equations (5), namely, the system of nonlinear $m$ equations for the $m$ tunable SRO parameters are taken. The equations are solved using a numerical root-finding algorithm such as the Newton-Raphson method[24]. The Newton-Raphson update step (Step 3 in Figure 3a) requires the flux and Jacobian associated with $\mathbf{z}$. At the end, the success of the exercise hinges on the flux of all 11 SRO parameters becoming zero at $\mathbf{z}_{eq}$.

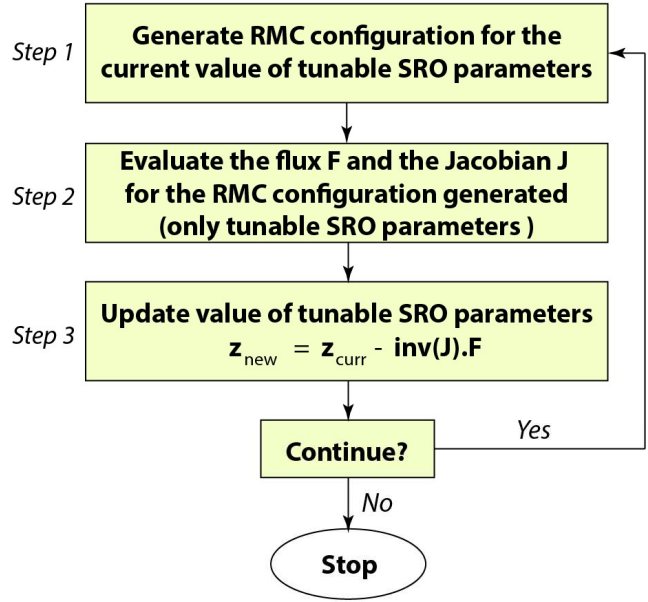

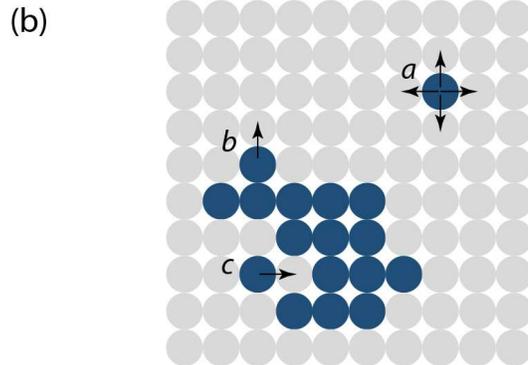

Figure 3: (a) Flowchart for identifying $z_{eq}$ using RMC. (b) Hop moves for the calculation of flux for current value of $z$.

The flux $F_{cluster}(z)$ associated with a particular type of cluster ($1NN$, $2NN$, etc.) is calculated in the following manner. Consider an occupied site $s$. A hop move is attempted to a 1NN vacant site $s'$ with a rate

$$r_{s \to s'} = \min(1, \exp\{-\beta \Delta U_{s \to s'}\}).  \quad (6)$$

Such a form is chosen taking into account thermodynamic considerations[23]. Figure 3b shows examples of such moves using the arrow symbol. A move may alter the local arrangement of atoms. $\Delta U_{s \to s'}$ is the associated change in energy, which is calculated using Equation (1). $\beta = (k_B T)^{-1}$, $k_B$ is the Boltzmann constant and $T$ is the temperature. The local change in the cluster count $v_{s \to s'}^{cluster}$ is analogous to a stoichiometric coefficient for chemical reactions. For atom $a$ in Figure 3b, $v_{s \to s'}^{1NN}$ is zero in all four hop directions. On the other hand, $v_{s \to s'}^{1NN} < 0$ and $v_{s \to s'}^{1NN} > 0$ for atoms $b$ and $c$, respectively, for the directions shown in Figure 3b. Since the purpose is only to evaluate $r_{s \to s'}$ and $v_{s \to s'}^{cluster}$, the hopping atom is returned to its original position after this is done.

The overall flux for the current RMC configuration at $\mathbf{z}$ is

$$F_{cluster}(\mathbf{z}) = \sum_{s,s'} r_{s \to s'} \, v_{s \to s'}^{cluster}. \quad (7)$$

The summation is performed over all possible atom hop moves in the configuration. Equation (7), or the *flux equation*, lays the foundation for incorporating multiple SRO parameters. From Equation (7) it is clear that the rate depends on the interaction parameters and the temperature. As seen in Figure 1a-c, $z_{1NN}$ determines the probability distribution of local arrangements. The flux can be positive, negative or zero. When the flux is zero for all SRO parameters, a stationary distribution for the local arrangements is obtained. In this way, the overall tendency for the system to move away from the current configuration is measured.

In principle, one may include other types of moves as well. Analogous to a MC simulation, a particle swap move, wherein an $A - V$ pair is randomly selected and the position of the $A$ and $V$ are interchanged, is an example of a move that may be considered. However, in such a case, the summation in Equation (7) will contain $N_A N_V$ terms if all possible moves are to be considered. On the other hand, hopping to neighboring site as in Figure 3 results in order $O(N_A)$ terms making the book-keeping simpler. Particle insertion moves are used later for the calculation of the chemical potential, once $z_{eq}$ is available (see discussion in Section 4.1).

## 3. Simulation details

### 3.1 Reverse Monte Carlo

All RMC simulations are performed with a square 2D lattice containing more than $10^5$ sites. Periodic boundary conditions are applied. The RMC algorithm is identical to the one in Ref. 18. A RMC converged configuration is generated within a few seconds on a single computer processor[25]. The CPU time requirement increases with the number of tunable parameters. It should be mentioned that our present implementation for RMC is not fully optimized, and there is scope for reducing the computational overhead[26], if the need arises. To improve the statistical sampling of local arrangements, four independent RMC configurations are generated for each $z$ value using different random seeds. This allows local arrangements appearing with a low probability of $\sim 10^{-5}$ to be sampled. The flux reported in Section 4 is an average of the four configurations.

### 3.2 Solving for $z_{eq}$

The Newton-Raphson method[24] is an iterative procedure used here to obtain the roots of the flux equation (Figure 3a). The initial guess value for $\mathbf{z} = \mathbf{z_0}$ provided corresponds to one for a well-mixed arrangement. Let the calculated flux for the $m$ tunable SRO parameters be $\mathbf{F}(\mathbf{z})$. The Jacobian matrix $\boldsymbol{J}(\boldsymbol{z})$ is evaluated using a central difference scheme with a small step of $\Delta z = 0.01$. A backward or forward difference scheme can be employed at the boundary. The new value of the SRO parameters at iteration $p+1$ is determined from one at iteration $p$ using $\boldsymbol{z}_{p+1} = \boldsymbol{z}_p - \boldsymbol{J}(\boldsymbol{z}_p)^{-1}\boldsymbol{F}(\boldsymbol{z}_p)$. The SRO parameter is forced to the boundary in case the updated value is out of bounds.

### 3.3 Grand canonical Monte Carlo

Grand Canonical Monte Carlo (GCMC) simulations are performed for making comparisons to our RMC approach. In GCMC, one specifies the chemical potential $\mu$ and temperature $T$ to obtain the value of $x$. For Cl adsorption on Cu(100) surface, the gas phase chemical potential $\mu_{Cl}$ of a Cl atom and temperature $T = 77\ K$ is provided as an input. This is in contrast to the RMC approach where $x$ and $T$ are specified and $\mu$ is an output. The adsorption sites are arranged in the form of a 50×50 2D lattice (2500 sites). Compared to RMC, a smaller lattice is used because GCMC calculations entail a random walk in the configurational space making it computationally expensive. This is particularly true for Cl adsorption on Cu(100) surface in Section 4.3. As mentioned earlier, each site can be either empty or occupied by an atom. Periodic boundary conditions are used.

Three different trial moves[23], namely, swap, insertion and deletion are considered:

1. Swap: Randomly selected atom in the lattice is moved to another randomly selected vacant site with an acceptance probability

$$p_{swap} = \min(1, \exp(-\beta \Delta U)). \qquad (8)$$

2. Insertion: An atom is inserted at a randomly selected vacant site with an acceptance probability

$$p_{ins} = \min\left(1, \frac{N_t - N_A}{N_A + 1} \exp(\beta\mu_A - \beta\Delta U)\right). \quad (9)$$

3. Deletion: A randomly selected atom is removed from the lattice with an acceptance probability

$$p_{del} = \min\left(1, \frac{N_A}{N_t - N_A + 1} \exp-(\beta\mu_A + \beta\Delta U)\right) \quad (10)$$

Here, $N_t$ is the total number of lattice sites, $N_A$ is the number of filled sites, $\Delta U$ is the energy change (energy for the new configuration minus the old one when a move is attempted). The system energy is calculated using a cluster expansion model Equation (1). Test for equilibrium involves the number of adsorbed atoms and the system energy becoming constant.

## 4. Results and discussion

Three examples, each with a different set of interactions, are considered (see Table 2). The first example uses a single tunable SRO parameter to validate our approach. The results are compared to our previous implementation of RMC, which also uses a single SRO parameter. The second example illustrates the use of multiple tunable SRO parameters. Finally, the third example involves a real application, namely, chemisorption of chlorine on copper.

**Table 2. Dimensionless cluster interactions used in different examples. See Figure 2 for meaning of the cluster. For Cl adsorption, T=77 K.**

| Cluster interaction | Example 1<br>One tunable SRO parameter | Example 2<br>Multiple tunable SRO parameter | Example 3<br>Cl adsorption on Cu(100) |
|---|---|---|---|
| $\beta w_{1NN}$ | -0.6 | -0.3 | 100 |
| $\beta w_{2NN}$ | 0 | -0.2 | 0.6458 |
| $\beta w_{3NN}$ | 0 | -0.1 | 0.0696 |
| $\beta w_{4NN}$ | 0 | 0 | 0.0077 |
| $\beta w_L$ | 0 | 0 | 0.4138 |
| $\beta w_I$ | 0 | 0 | 0.6303 |
| $\beta w_V$ | 0 | 0 | 0.3519 |
| $\beta w_\Delta$ | 0 | 0 | -0.1624 |
| $\beta w_\diamond$ | 0 | 0 | 0.0541 |
| $\beta w_Q$ | 0 | 0 | 0.0155 |
| $\beta w_\square$ | 0 | 0 | -0.1508 |

## 4.1 One tunable SRO parameter

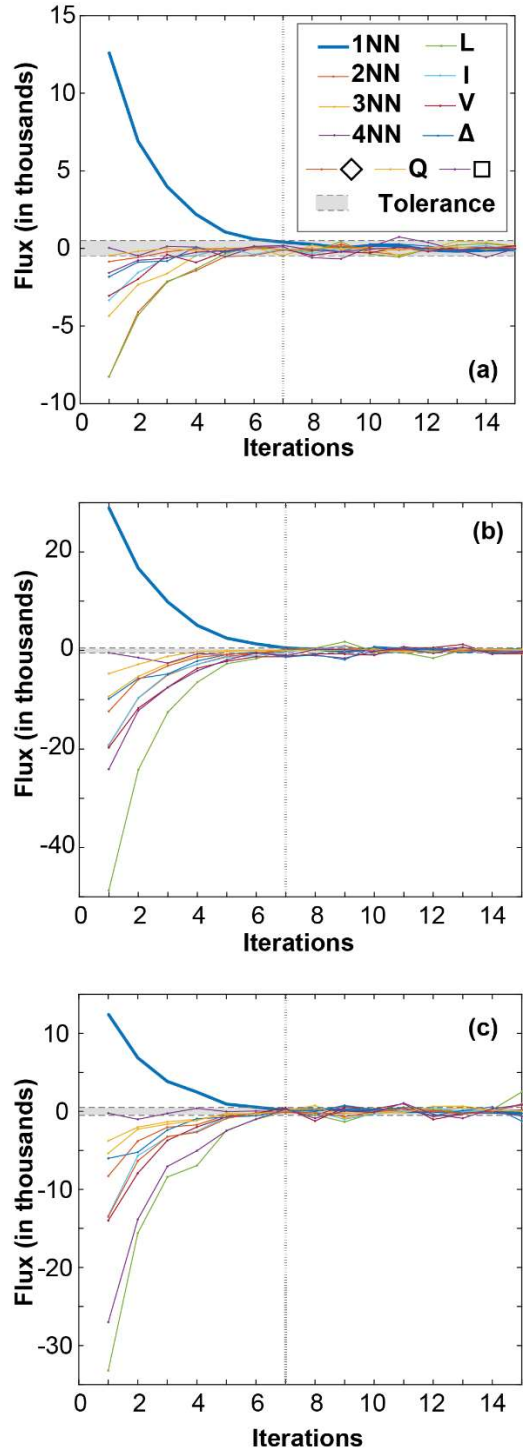

Figure 4: Flux calculated at different Newton-Raphson iterations when (a) $x = 0.2$, (b) $x = 0.5$, (c) $x = 0.8$.

The goal is to calculate the gas phase chemical potential $\mu_A$ as a function of $x$ when only one type of interaction $\beta w_{1N} = -0.6$ is present. RMC configurations are generated using one tunable SRO parameter, namely, $z_{1NN}$. The equilibrium $z_{1NN}$ is determined using flux measurement. Figure 4 demonstrates the flux convergence behavior for three different $x$ values (0.2, 0.5 and 0.8). The same behavior is observed in all cases. Initially, the flux associated with $z_{1NN}$ is large, in the order of $10^4$. Flux can be made dimensionless by dividing by the system size (here $N_t \approx 10^5$), or as mentioned earlier more appropriately by $bN_a$. The flux decreases to a value less than the set tolerance of 500. The tolerance value used is shown by the two dashed horizontal lines in Figure 4. The system is deemed as equilibrated once the magnitude of the flux for $z_{1NN}$ is within the tolerance. The vertical dotted line in each panel shows that only seven Newton-Raphson iterations are required to reach equilibrium. Thereafter, the flux values fluctuate around zero within the set tolerance. Flux for other SRO parameters is also evaluated. These exhibit a behavior similar to $1NN$. Their magnitudes also reduce and converge synchronously with the 1NN parameter. After convergence, the change in the SRO parameters due the fluctuations is in the order of $10^{-3}$, and thus negligible.

The converged value of $z_{1NN}$ is compared to our earlier study [19] in Figure 5. Excellent agreement is seen. We find that $z_{1NN} > x$ (Figure 5a) implying that the $A$ atoms tend to cluster because of the attractive pair interactions. The equilibrium configuration is shown for $x = 0.5$ in inset of Figure 5a.

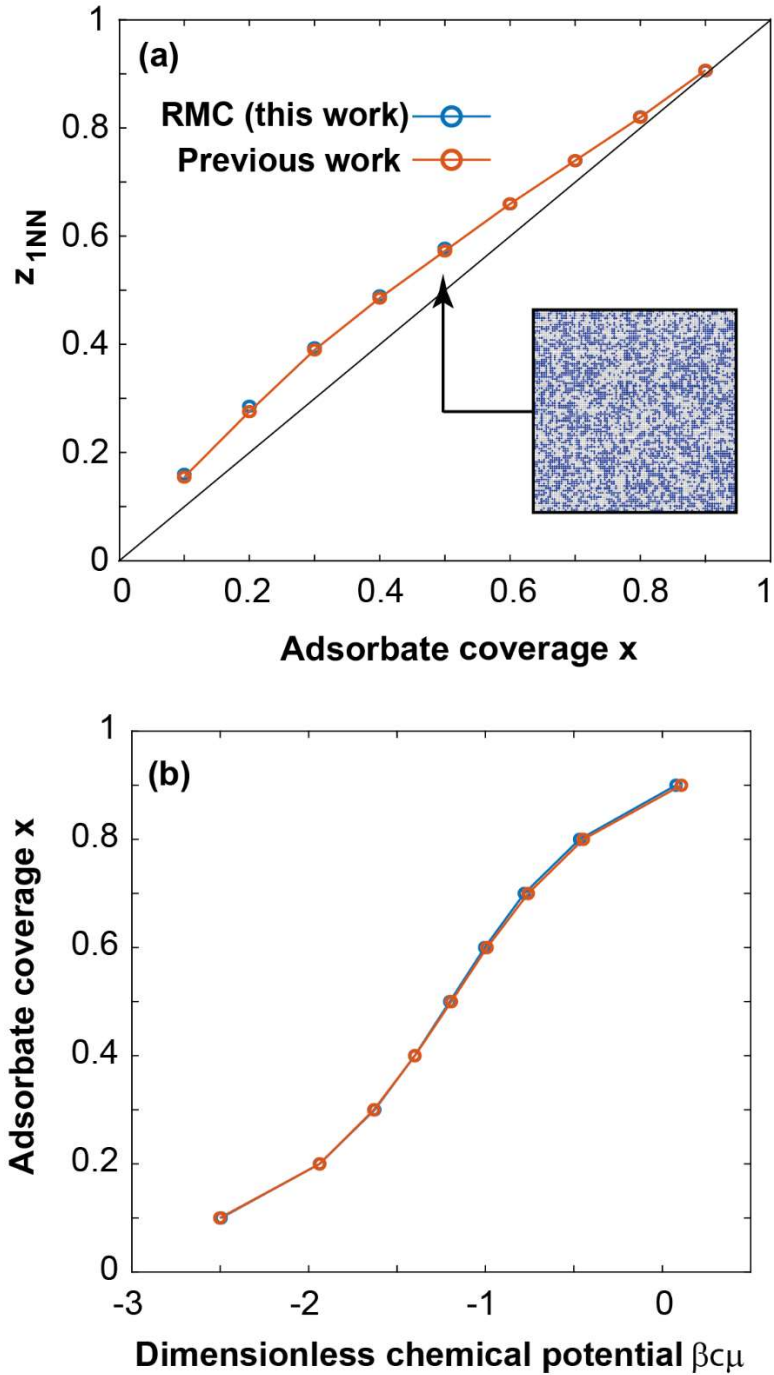

Figure 5: (a) $z_{1NN}$ obtained in terms of the coverage $x$ matches with our previous work [19]. (b) Corresponding adsorption isotherm, $c = 4$.

The chemical potential is evaluated using a particle insertion technique. In this approach, an $A$ atom is inserted in a vacant site of an equilibrated configuration only for the purpose of calculating the energy change $\Delta U$. The excess (gas phase) chemical potential is determined using

$$\mu_A = \ln\langle \exp(-\beta \Delta U)\rangle. \tag{11}$$

Figure 5b shows that the equilibrium 1NN SRO parameter and the chemical potential of the final configuration are in good agreement with the previous results as well as GCMC (GCMC results are not shown) [19]. Adding more tunable parameters does not change the results. This indicates that using a single tunable SRO parameter is adequate.

### 4.2 Three tunable SRO parameters

An example is discussed where multiple SRO parameters become necessary. Table 2 provides the interaction strengths. Negative values indicate that the $A - A$ interactions are attractive and cluster formation is preferred. First, we tune only one SRO parameter, namely $z_{1NN}$. Figure 6a shows the convergence behavior of the flux for different SRO parameters. As in Section 4.1, the flux for $1NN$ is initially large in magnitude but it decreases and approaches zero within 6 iterations. However, the behavior for other clusters is not the same. For e.g., the flux for $2NN$ increases and subsequently reaches a steady value, which is much greater than the set tolerance. The tolerance value used is shown by the two dashed lines in Figure 6a. This suggests that the 2D configuration is in fact unstable. Since the $2NN$ local arrangements are not being tuned, we have no control over the associated flux. A steady value of the flux is reached only because $z_{1N}$ becomes practically constant after 6 iterations.

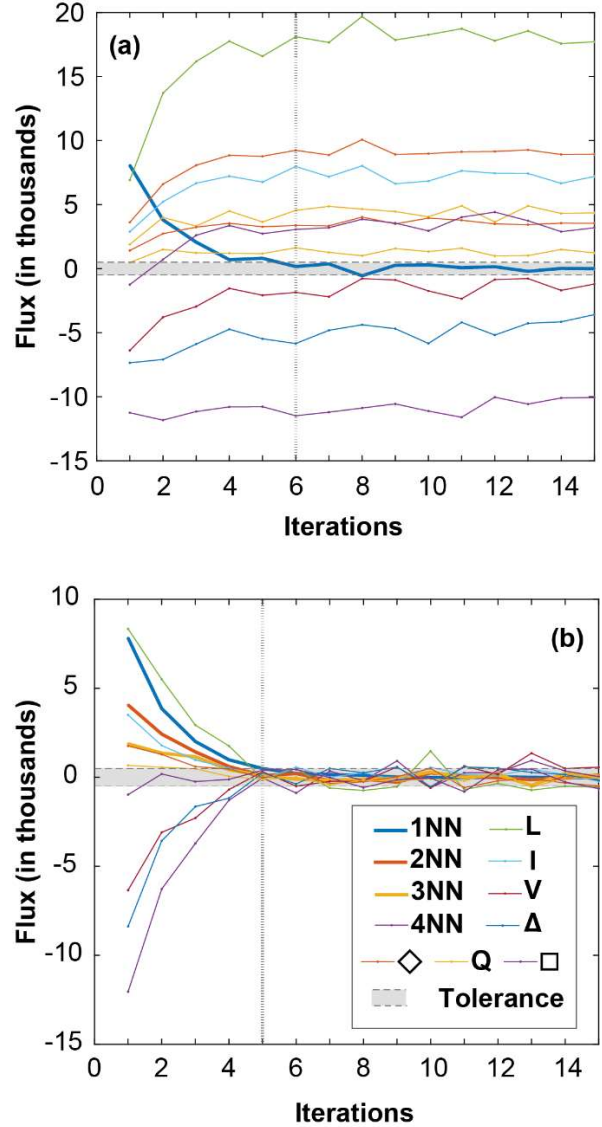

Figure 6: Bold lines indicate the flux for the tuned SRO parameters. Flux is plotted in terms of the Newton-Raphson iteration. (a) Only $1NN$ and (b) $1NN$, $2NN$ and $3NN$ are tuned. Here $x = 0.4$.

To remedy this, we incorporate more SRO parameters into our analysis. Results are discussed for the case where we tune three SRO parameters, namely $z_{1NN}$, $z_{2NN}$ and $z_{3NN}$. It is interesting to note from Figure 6b that the flux for all SRO parameters now synchronously approaches zero.

This suggests that various local pair, triplet and quadruplet arrangements are stable. Equilibrium is reached within a specified tolerance is obtained after 5 iterations. The fluctuations in the flux beyond 5 iterations are due to the statistical noise inherent in RMC simulations. As in the first example, the flux reported is obtained by averaging over 4 independent RMC configurations for the given $z$. The fluctuations can be made smaller with more averaging. Figure 7 shows a comparison of the adsorption isotherms. It is observed that RMC and GCMC are in good agreement.

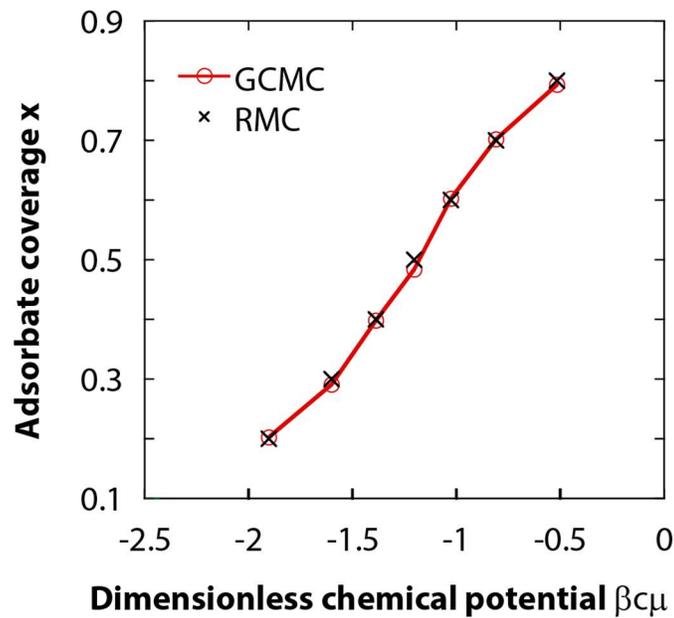

Figure 7: Adsorption isotherm obtained using RMC (with 3 tunable SRO parameters) and GCMC. Here $c = 4$.

Another aspect of thermodynamic behavior is the deviation from ideality. Traditionally, deviation is measured in terms of an excess property. However, from a structure-property relation point-of-view one can enquire about the deviation structurally. Table 1 provides the expression for the SRO parameters for the well mixed arrangement. We assess the deviation from ideality as $\ln(z_{cluster}/z_{cluster}^{well-mixed})$. The deviation is zero for the well mixed arrangement. In such a case, the excess property also becomes zero. Figure 8 shows the deviation for different SRO parameters as a function of $x$. The first observation is that systematic trends are obtained, which suggest that the statistical noise in the RMC data is small. In contrast, GCMC results are found to be less systematic and quite noisy (not shown). We also observe that certain SRO parameters such as $4NN$ behave more ideally than others (e.g., $1NN$) over the entire range of $x$. Quadruplet Q1 and the L triplet possess the largest deviation from ideality. All SRO parameters behave ideally for $x > 0.9$. Thus, one can expect mean-field theory to be valid when $x > 0.9$.

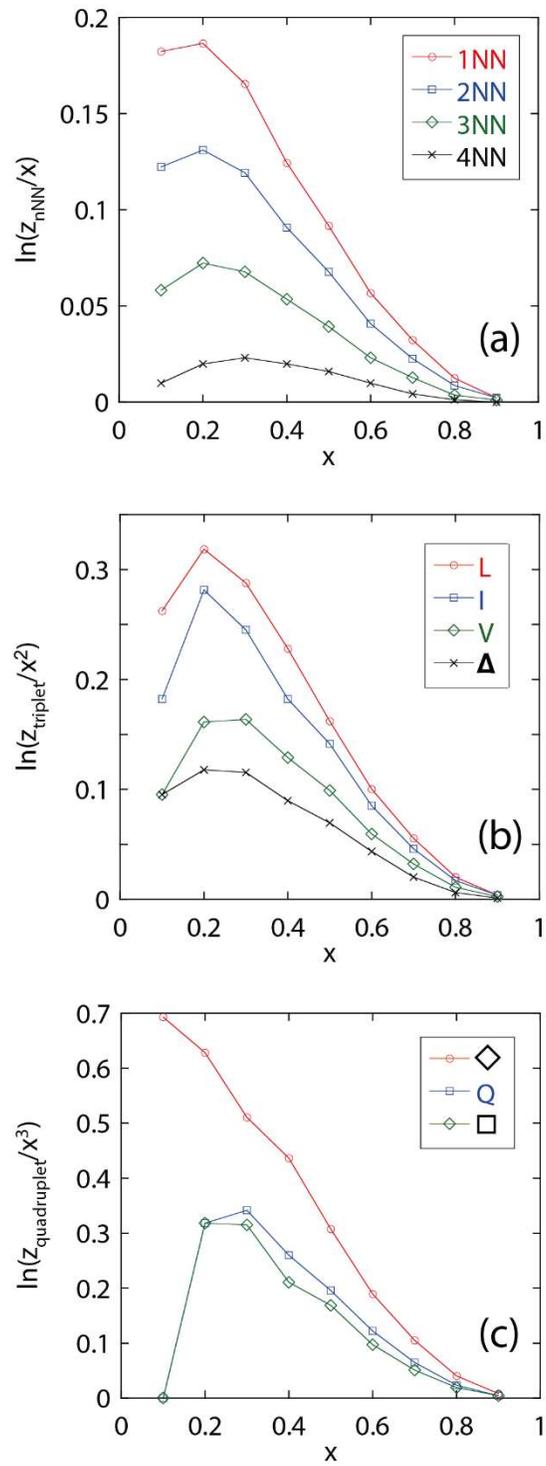

Figure 8. Deviation from ideality observed with the help of RMC using 3 tunable SRO parameters.

**4.3 Cl adsorption on Cu(100)**

Finally, the RMC approach is applied to a real system, specifically for *ab initio* thermodynamics calculations involving the chemisorption of chlorine on a copper(100) surface. Adsorbed halides are known to modify the properties of transition metal catalysts due to their strong interaction with metals. STM image shows that Cl adsorbs on Cu forming an ordered c(2×2) adlayer arrangement [27]. Figure 9b shows such an arrangement. Given its significance, it is desirable that such arrangements can be modelled using theory.

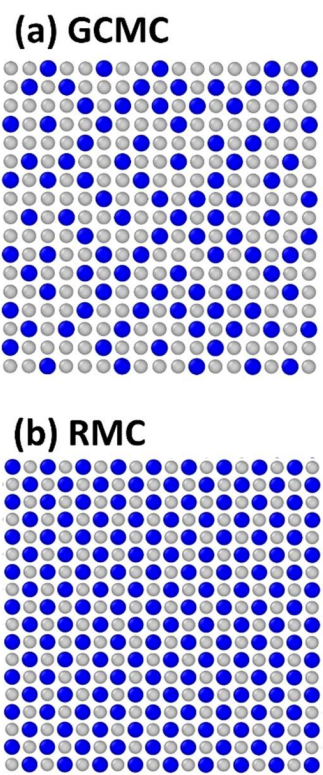

Figure 9: Comparison of the final Cl adlayer configuration on Cu(100) obtained from (a) GCMC and (b) RMC calculations using the same adsorbate-adsorbate interactions at high values of Cl chemical potentials. Only the RMC configuration (with Cl coverage $x = 0.5$) is found to be identical to the one observed experimentally.

The Cl-Cl cluster interactions are given in Table 2. The energy of the chlorine adlayer system (with respect to the reference state containing bare Cu slab and $n_{Cl}$ Cl atoms in gas phase) is obtained as

$$E_{system} = n_{Cl}E_{ad}^0 + E_{adsorbate\ interactions}. \qquad (12)$$

In the dilute limit, the adsorption energy of Cl on Cu(100) is found to be $E_{ad}^0 = -2.4579\ eV$ using density functional theory calculations [28]. At higher chlorine coverage, the adsorption energy decreases due to many-body lateral repulsive interactions between the Cl atoms. The adsorbate-adsorbate interactions are calculated using a cluster expansion model

$$E_{adsorbate\ interactions} = \sum_{cluster} w_{cluster} n_{cluster}. \qquad (13)$$

This CEM was trained using a DFT dataset, which consisted of the calculated energy ($E$) and the cluster counts ($n_{cluster}$) for 42 different Cl coverages/configuration. The effective cluster interactions (ECIs), denoted as $w_{cluster}$ in Equation (13), are given in Table 2. Details of the training procedure can be found in Ref. [28]. Here we employ the CEM with our GCMC and RMC calculations. Unlike the previous two examples, here most of the interactions are positive, i.e., repulsive (Table 2).

GCMC simulation was performed starting with a bare Cu surface. At high chemical potential, the GCMC simulations reach a maximum coverage of ~$0.36\ ML$, which is much lower than the $0.5\ ML$ coverage expected with a c(2×2) arrangement. This can be seen in Figure 9a where a sufficiently large $\mu_{Cl} = -2\ eV$ is provided (see Ref. [28]). The coverage remained unchanged even after several days of running the simulation. In most circumstances, the interaction model will be blamed for the failure. However, the main challenge in GCMC is that a c(2×2) arrangement has a very low probability of being selected.

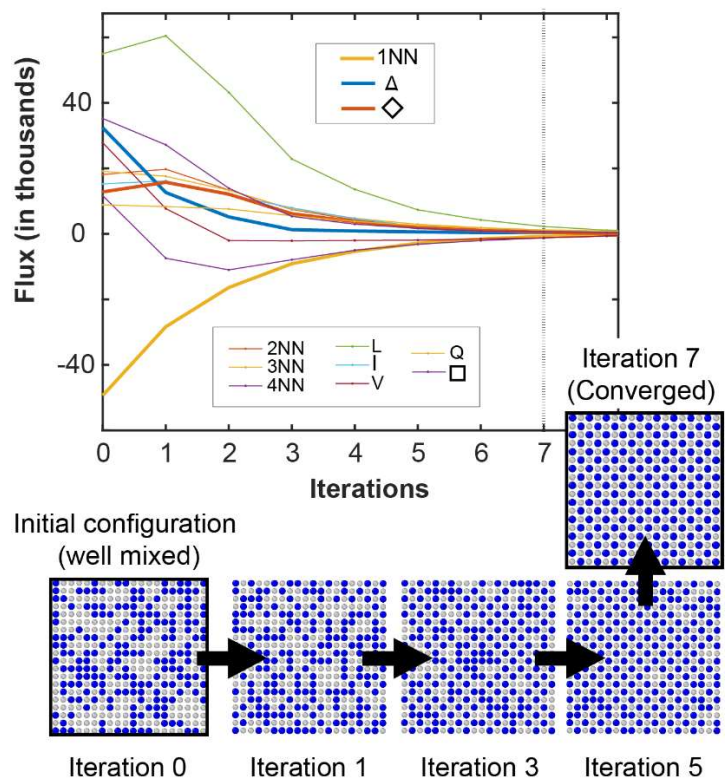

Figure 10: Flux as a function of Newton-Raphson iterations as the c(2×2) adlayer arrangement of chlorine on Cu(100) finally emerges. Starting with the initial perfectly well-mixed arrangement, a set of intermediate arrangements to finally the converged (equilibrium) arrangement is shown.

For RMC we tune 3 SRO parameters, namely $z_{1N}$, $z_\Delta$ and $z_\diamond$. The choice of these three clusters is made to show that (i) SRO parameters other than pairs can also be tuned and (ii) three parameters are enough to achieve the full ordered structure even though there are several non-zero interaction terms in Table 2. Whether any another combination of SRO parameters would have given adequate results is beyond the scope of the present work. A well-mixed arrangement with $x = 0.5$ is provided as an initial guess to the Newton-Raphson algorithm, Figure 10 shows that the flux becomes zero after 7 iterations. The corresponding evolution of the surface is also shown at iteration 0, 1, 3, 5 and 7. Figure 9b shows a part of the equilibrated RMC lattice more clearly. Table 3 shows the SRO parameter at equilibrium from the different methods. The configuration generated by RMC does contain a few defects. For example, adsorbed Cl is present in few places at the 1NN and 4NN positions. As a result, $z_1$ and $z_\Delta$ in Table 3 are not zero and $z_{Q1}$ is less than one. For a perfect c(2×2) adlayer, $z_{1NN}$ and $z_\Delta$ should be zero, whereas $z_{Q1}$ should be one. Defects in the c(2×2) adlayer have not been investigated thoroughly in literature. It should be noted that when the RMC configuration was provided to GCMC it was found to be stable. In conclusion, for ordered adlayer structures RMC performs significantly better than GCMC in reducing imperfections and defects in the generated configurations.

Table 3. Maximum coverage observed with Cl/Cu(100) and associated SRO parameters.

| Method | x | 1NN | 2NN | 3NN | 4NN | L1 | I | L2 | L3 | Q1 | Q2 | Q3 |
|---|---|---|---|---|---|---|---|---|---|---|---|---|
| RMC | 0.5 | 0.065 | 0.890 | 0.870 | 0.159 | 0.825 | 0.7960 | 0.105 | 0.092 | 0.780 | 0.184 | 0.366 |
| GCMC | 0.349 | 0.000 | 0.580 | 0.457 | 0.287 | 0.336 | 0.346 | 0.155 | 0.120 | 0.248 | 0.116 | 0.177 |
| Experiment | 0.5 | 0 | 1 | 1 | 0 | 1 | 1 | 0 | 0 | 1 | 0 | 1 |

## 5. Conclusion

We have introduced a computational strategy that involves use of the reverse Monte Carlo (RMC) method for lattice thermodynamic property calculations. On-lattice structures are generally encountered in crystalline solids, surface science, alloys, adsorption, absorption, separation, catalysis, etc., making them an important class of problems[21]. The main highlight is the new computational framework that allows for simultaneous tuning of multiple short-range order (SRO) parameters. Previous variations of the RMC-based thermodynamic calculations were limited in this respect as they could tune only one SRO parameters. The use of multiple SRO parameters provides access to a much wider configurational space. It has been shown here that multiple SRO parameters are required when longer-ranged interactions and ordered structures are present.

In the new approach, a system of nonlinear algebraic flux equations is solved to obtain the SRO parameters at equilibrium. These equations can be solved using standard numerical techniques, such as Newton Raphson method. An important observation is that while grand canonical Monte Carlo (GCMC) simulations often sample several millions of configurations, the RMC method can reach the equilibrium configuration within ten Newton Raphson iterations (or few tens of configurations sampled). This makes the RMC method more computationally efficient. In two examples, RMC and GCMC were found to be in excellent agreement. It should be noted that in the third example, GCMC was unable to provide the correct ordered chlorine adsorbed layer structure on copper (100) surface even after several days of calculation, whereas RMC required only 7 iterations. Similar performance is expected for 3D lattice problems.

Another observation is that in principle one may specify/monitor a very large number of SRO parameters. However, out of these only a few parameters are found to be independent. The independent SRO parameters control the behavior of others. When the independent SRO parameters reach an equilibrium value, so do others. We have investigated the behavior of 11

SRO parameters in this paper. The number of "independent" SRO parameters varied between 1-3. This can be exploited to simplify the RMC implementation.

This paper introduces several directions which can be taken in future. First, a more systematic procedure to identify the independent SRO parameters is needed. Second, the present study is restricted to a two-component system. Extension to multiple species and multiple binding sites should be attempted. A preliminary study suggests that the number of SRO parameters may scale linearly with the number of species[16]. Finally, the RMC method has recently been also applied to nonequilibrium problems while employing one tunable SRO parameter [26]. Based on the findings here, extension to multiple tunable SRO parameters (for example, analogous to Equation (4)) will be desirable.

**Appendix**

Figure 1(d)-(f) are generated using RMC with the following parameters:

1(d): $z_{1NN}$ = 0.640, $z_{2NN}$ = 0.536, $z_{3NN}$ = 0.329, $z_{4NN}$ = 0.428

1(e): $z_{1NN}$ = 0.500, $z_{2NN}$ = 0.527, $z_{3NN}$ = 0.528, $z_{4NN}$ = 0.736

1(f): $z_{1NN}$ = 0.320, $z_{2NN}$ = 0.320, $z_I$ (only left diagonal) = 0.428.

**Acknowledgements**

The authors acknowledge Mr. Bibek Dash (Institute of Minerals and Materials Technology, Bhubaneswar) for performing DFT simulations of Cl/Cl(100) system, which were utilized in calculating the configurational free energy of the Cl adlayer.

AC acknowledges support from Science and Engineering Research Board Grant No. EMR/2017/001520, CRG/2022/008058 and MTR/2019/000909 and National Supercomputing Mission DST/NSM/R&D_HPC_Applications/2021/02.**References**

(1) McGreevy, R. L.; Pusztai, L. Reverse Monte Carlo Simulation: A New Technique for the Determination of Disordered Structures. *Mol. Simul.* **1988**, *1* (6), 359–367. https://doi.org/10.1080/08927028808080958.

(2) McGreevy, R. L.; Howe, M. A. RMC: Modeling Disordered Structures. *Annu. Rev. Mater. Sci.* **1992**, *22* (1), 217–242. https://doi.org/10.1146/annurev.ms.22.080192.001245.

(3) Wikfeldt, K. T.; Leetmaa, M.; Ljungberg, M. P.; Nilsson, A.; Pettersson, L. G. M. On the Range of Water Structure Models Compatible with X-Ray and Neutron Diffraction Data. *J. Phys. Chem. B* **2009**, *113* (18), 6246. https://doi.org/10.1021/jp9007619.

(4) Harsányi, I.; Pusztai, L. Hydration Structure in Concentrated Aqueous Lithium Chloride Solutions: A Reverse Monte Carlo Based Combination of Molecular Dynamics Simulations and Diffraction Data. *J. Chem. Phys* **2012**, *137*, 204503. https://doi.org/10.1063/1.4767437.

(5) Tucker, M. G.; Keen, D. A.; Dove, M. T.; Goodwin, A. L.; Hui, Q. RMCProfile: Reverse Monte Carlo for Polycrystalline Materials. *J. Phys. Condens. Matter* **2007**, *19* (33). https://doi.org/10.1088/0953-8984/19/33/335218.

(6) Keen, D. A.; McGreevy, R. L. Determination of Disordered Magnetic Structures by RMC Modelling of Neutron Diffraction Data. *J. Phys. Condens. Matter* **1991**, *3* (38), 7383–7394.31

https://doi.org/10.1088/0953-8984/3/38/010.

(7) Gurman, S. J.; McGreevy, R. L. Reverse Monte Carlo Simulation for the Analysis of EXAFS Data. *J. Phys. Condens. Matter* **1990**, *2* (48), 9463–9473. https://doi.org/10.1088/0953-8984/2/48/001.

(8) Howe, M. A.; Mcgreevy, R. L.; Pusztai, L.; Borzsák, & I. Determination of Three Body Correlations in Simple Liquids by RMC Modelling of Diffraction Data. *Phys. Chem. Liq.* **1993**, *25* (4), 205–241. https://doi.org/10.1080/00319109308030363.

(9) Veglio, N.; Bermejo, F. J.; Pardo, L. C.; Ll Tamarit, J.; Cuello, G. J. Direct Experimental Assessment of the Strength of Orientational Correlations in Polar Liquids. *Phys. Rev. E* **2005**, *72*, 031502. https://doi.org/10.1103/PhysRevE.72.031502.

(10) McGreevy. R.L.; Pusztai L. The Structure of Molten Salts. *Proc. R. Soc. London. Ser. A Math. Phys. Sci.* **1990**, *430* (1878), 241–261. https://doi.org/10.1098/RSPA.1990.0090.

(11) Kaban, I.; Jóvári, P.; Stoica, M.; Eckert, J.; Hoyer, W.; Beuneu, B. Topological and Chemical Ordering in Co 43 Fe 20 Ta 5.5 B 31.5 Metallic Glass. *Phys. Rev. B* **2009**, *79*, 212201. https://doi.org/10.1103/PhysRevB.79.212201.

(12) Keen, D. A.; McGreevy, R. L. Structural Modelling of Glasses Using Reverse Monte Carlo Simulation. *Nature* **1990**, *344* (6265), 423. https://doi.org/10.1038/344423a0.

(13) Keen, D. A.; Tucker, M. G.; Dove, M. T. Reverse Monte Carlo Modelling of Crystalline Disorder. *J. Phys. Condens. Matter* **2005**, *17* (5), S15–S22. https://doi.org/10.1088/0953-8984/17/5/002.

(14) Thomson, K. T.; Gubbins, K. E. Modeling Structural Morphology of Microporous Carbons by Reverse Monte Carlo. *Langmuir* **2000**, *16* (13), 5761–5773.

https://doi.org/10.1021/la991581c.

(15) Pikunic, J.; Clinard, C.; Cohaut, N.; Gubbins, K. E.; Guet, J. M.; Pellenq, R. J. M.; Rannou, I.; Rouzaud, J. N. Structural Modeling of Porous Carbons: Constrained Reverse Monte Carlo Method. *Langmuir* **2003**, *19* (20), 8565–8582. https://doi.org/10.1021/la034595y.

(16) Agrahari, G.; Chatterjee, A. Speed-up of Monte Carlo Simulations by Preparing Starting off-Lattice Structures That Are Close to Equilibrium. *J. Chem. Phys.* **2020**, *152* (4), 44102. https://doi.org/10.1063/1.5131303.

(17) Agrahari, G.; Chatterjee, A. Thermodynamic Calculations Using Reverse Monte Carlo. *Phys. Rev. E* **2021**, *104*, 044129.

(18) Agrahari, G.; Chatterjee, A. Thermodynamic Calculations Using Reverse Monte Carlo: Convergence Aspects, Sources of Error and Guidelines for Improving Accuracy. *Mol. Simul.* **2022**, *48* (13), 1143–1154.

(19) Ball, A. K.; Swati, R.; Agrahari, G.; Chatterjee, A. Accelerated Calculation of Configurational Free Energy Using a Combination of Reverse Monte Carlo and Neural Network Models: Adsorption Isotherm for 2D Square and Triangular Lattices. *Comput. Phys. Commun.* **2023**, *285*, 108654. https://doi.org/doi.org/10.1016/j.cpc.2022.108654.

(20) Porter, D. A.; Easterling, K. E.; Sherif, M. *Phase Transformations in Metals and Alloys*; CRC Press: Boca Raton, 2009.

(21) Hill, T. L. *An Introduction to Statistical Thermodynamics*; Dover: New York, 1986.

(22) Sandler, S. I. *Chemical and Engineering Thermodynamics*, 3rd ed.; John Wiley & Sons: New York, 1999.

(23) Frenkel, D.; Smit, B. *Understanding Molecular Simulation: From Algorithms to Applications*;

Academic Press: New York, 1996.

(24) Press, W. H.; Flannery, B. P.; Teukolsky, S. A.; Vetterling, W. T. *Numerical Recipes*; Cambridge University Press: Cambridge, 1986.

(25) Ball, A. K.; Haque, S.; Chatterjee, A. Relaxation Dynamics in Lattice Reverse Monte Carlo. *Mol. Simul.* **2023**, 1–13. https://doi.org/10.1080/08927022.2023.2202780.

(26) Kumar, A.; Chatterjee, A. A Probabilistic Microkinetic Modeling Framework for Catalytic Surface Reactions. *J. Chem. Phys.* **2023**, *158*, 024109. https://doi.org/doi.org/10.1063/5.0132877.

(27) Suggs, D. W.; Bard, A. J. Scanning Tunneling Microscopic Study with Atomic Resolution of the Dissolution of Cu(100) Electrodes in Aqueous Chloride Media. *J. Phys. Chem.* **1995**, *99* (20), 8349–8355. https://doi.org/10.1021/j100020a070.

(28) Dash, B.; Haque, S.; Chatterjee, A. Reduced Collinearity, Low-Dimensional Cluster Expansion Model for Adsorption of Halides (Cl, Br) on Cu(100) Surface Using Principal Component Analysis. *Submitted* **2023**.